\documentclass[12pt]{article}

\begin{document}


\title{Upper Limit Set by Causality on the Rotation and Mass of Uniformly 
Rotating Relativistic Stars}

\author{Scott Koranda \\
Department of Physics, Case Western Reserve University,\\
10900 Euclid Avenue, Cleveland, OH 44106-7079\\
\\
Nikolaos Stergioulas\\
Department of Physics, University of Wisconsin-Milwaukee,\\
P.O. Box 413, Milwaukee, WI  53201\\
\\
John L. Friedman\\
Department of Physics, University of Wisconsin-Milwaukee,\\
P.O. Box 413, Milwaukee, WI  53201}
\date{\today}

\maketitle

\baselineskip 24 pt
\begin{abstract}
\baselineskip 24 pt
Causality alone suffices to set a lower bound on the period of
rotation of relativistic stars as a function of their maximum observed
mass.  That is, by assuming a one-parameter equation of state (EOS)
that satisfies $v_{\rm sound}<c$ and that allows stars with masses as
large as the largest observed neutron-star mass, $M_{sph}^{max}$, we
find $P[ms] > 0.282 + 0.196\ \left( M_{sph}^{max}/M_\odot-1.442
\right)$.  The limit does {\it not} assume that the EOS agrees with a
known low-density form for ordinary matter, but if one adds that
assumption, the minimum period is raised by a few percent.  Thus
the {\it current} minimum period of uniformly rotating stars, set by
causality, is 0.28ms (0.29ms for stars with normal crust). The minimizing
EOS yields models with a maximally soft exterior supported by a
maximally stiff core.  An analogous upper limit set by causality on
the maximum mass of rotating neutron stars requires a low-density
match and the limit depends on the matching density, $\epsilon_m$.  We
recompute it, obtaining a slightly revised value, $M_{rot}^{max}
\simeq 6.1 \left ( 2
\times 10^{14} \ { g/cm^3}/ \epsilon_m \right )^{1/2} \ \ {M_\odot}$.
\end{abstract}


\newpage

\section{INTRODUCTION}

The upper limit set by gravity on the rotation of neutron stars is          
sensitive to the equation of state (EOS) above nuclear density,  and  the
current large uncertainty in the EOS implies a large uncertainty in
the maximum rate of rotation.  One can, however, find an upper limit on
rotation which is independent of the EOS.  This limit is set by
causality together with the requirement that the EOS allows stars with
masses as large as the largest observed neutron-star mass.

Glendenning (1992) first estimated this causally limited 
period for gravitationally bound stars.  He considered a flexible
two-parameter, gamma-law ansatz for the EOS at high energy density; by
varying the two parameters, he found equations of state (EOSs) which
appeared to minimize the period as a function of the maximum
nonrotating (or spherical) mass $M^{max}_{sph}$ allowed by the EOS.
Our work improves upon his estimate in several ways.  First, where
Glendenning used an empirical formula to estimate the minimum period
{}from a set of nonrotating models, we construct models using a code 
that constructs rapidly rotating relativistic stars.
Second, our numerical investigation
singles out a simple analytic form for the EOS which minimizes the
period for a given $M^{max}_{sph}$.  This minimum-period EOS has a simple
physical interpretation; it allows the most centrally condensed star
while still supporting a mass $M^{max}_{sph}$.  Specifically, we find
that if $M^{ max}_{sph} =$ 1.442 ${M}_{\odot}$ (currently the largest
accurately measured mass of neutron stars
(Taylor \& Weisberg 1989), the minimum period allowed is 0.28ms. This minimum 
period is about 13 \% less than that estimated by
Glendenning (1992). Third, we emphasize that there exists
an upper limit on rotation set only by causality, which requires no 
matching to a low density EOS. Finally, we find
an exact scaling between the minimum period $P_{min}$ and $M^{ max}_{
sph}$.  The scaling is due to the specific form of the minimum-period EOS,
and implies that the central energy densities, masses, and radii of
the maximum-mass spherical (and maximum-mass rotating) stars scale
between different minimum-period EOSs.

The maximum mass of gravitationally bound stars set by causality is, 
in contrast, quite sensitive to the matching density
(Hartle \& Sabbadini 1977) (the initial Rhoades and Ruffini (1974) limit
was computed for a particular choice of matching
density).  For rotating neutron stars, the maximum mass set by
causality was first obtained by Friedman and Ipser (1987).
This maximum mass has not, however, 
however been recomputed since recent improvements
in relativistic, rotating star codes, and we do so here.  Using the FPS EOS
(Pandharipande \& Ravenhall 1989)
at low density, we find $M_{ max} = 6.1 { M}_\odot
(\epsilon_m/2 \times 10^{14} { g/cm^3})^{-1/2}$, where $\epsilon_m$ is
the matching energy density.

\section{ASSUPMTIONS FOR THE EQUATION OF STATE}
\label{sec:eosconstraints}

For number densities $n\leq 0.1 \ { fm}^{-3}$ (slightly less than
nuclear saturation density) the FPS EOS 
(Pandharipande \& Ravenhall 1989) is expected to
describe normal matter with reasonable accuracy (Ravenhall 1995).
Above this density, however, current observations are compatible with
a variety of proposed equations of state that span a large range of
compressibilities and predict correspondingly different minimum
periods.  One can, however, use causality to set a lower limit on the
rotation period of stable relativistic stars that is independent of
the high density EOS.  More precisely, we assume:

\begin{itemize} 
\item[(0)] {\it A relativistic star is described as a self-gravitating,
uniformly rotating perfect fluid with a one-parameter EOS.}

Cold neutron stars satisfy this assumption to high accuracy 
(Friedman \& Ipser 1992).

\item[(1)] $v_{sound}\equiv\sqrt{dp/d\epsilon} \leq 1$.

Here $v_{sound}$ is the phase velocity of sound waves, $p$ the
pressure, and $\epsilon$ the energy density of the stellar perfect
fluid matter.  Relativistic fluids are governed by hyperbolic equations
whose characteristics lie outside the light cone (and thus their
initial value formulation violates causality) if $v_{ sound} > 1$ (here
and throughout the paper we have set 
$c=1$) (Geroch \& Lindblom 1991).  That is, if both the equilibrium star
and its small oscillations are described by a one-parameter EOS, then
assumption (1) is implied by causality.
\end{itemize}

\noindent These two assumptions alone yield an upper limit on 
rotation as a function of the maximum observed mass of a relativistic star.
Only a slightly more stringent limit is obtained if one makes the  
further asumption:

\begin{itemize} 
\item[(2)] {\it The EOS is known (matches FPS) at low density.}

Relativistic stars with normal crusts are thought to be accurately
described by the FPS EOS up to number densities $n\approx 0.1 \> {
fm}^{-3}$.  We adopt $n_m=0.1\> { fm}^{-3}$ as a conservative value for
the number density $n_m$ at which the EOS at high density must match
the FPS EOS for low densities.
\end{itemize}
 
\noindent  Below we examine limits on the rotation of stable
equilibrium models corresponding first to assumptions (0) and (1) only,
and then to assumptions (0)-(2); as noted, we find that the additional
assumption of a match to the known low-density equation of state has a
small effect on the minimum period computed using (0) and (1) only.

\section{EQUATION OF STATE YIELDING THE \\ UPPER LIMIT ON ROTATION}
\label{sec:conjecture}

A uniformly rotating, gravitationally bound star rotates with a period
greater than or equal to its Kepler period. The Kepler period is the
period of a free particle in circular orbit at the star's equator.  A
soft EOS yields stellar models with dense central cores, large binding
energies, and thus smaller Kepler periods than models built from stiff
EOSs.  Soft EOSs, however, cannot support massive stars.  This
suggests that models with minimum periods arise from EOSs which are
stiff at high density, allowing stiff cores to support against
collapse, but soft at low density, allowing small radii and thus fast 
rotation.
Our numerical results are consistent with this expectation, and
strongly suggest a simple form for the EOS yielding  minimum
period stars and satisfying  the assumptions listed in section
\ref{sec:eosconstraints}. We distinguish the two instances where
the EOS does or does not match smoothly to the FPS EOS at low
density and satisfy assumption $(2)$ above.

\subsection{Minimum-Period EOS}
\label{s:mincausal}

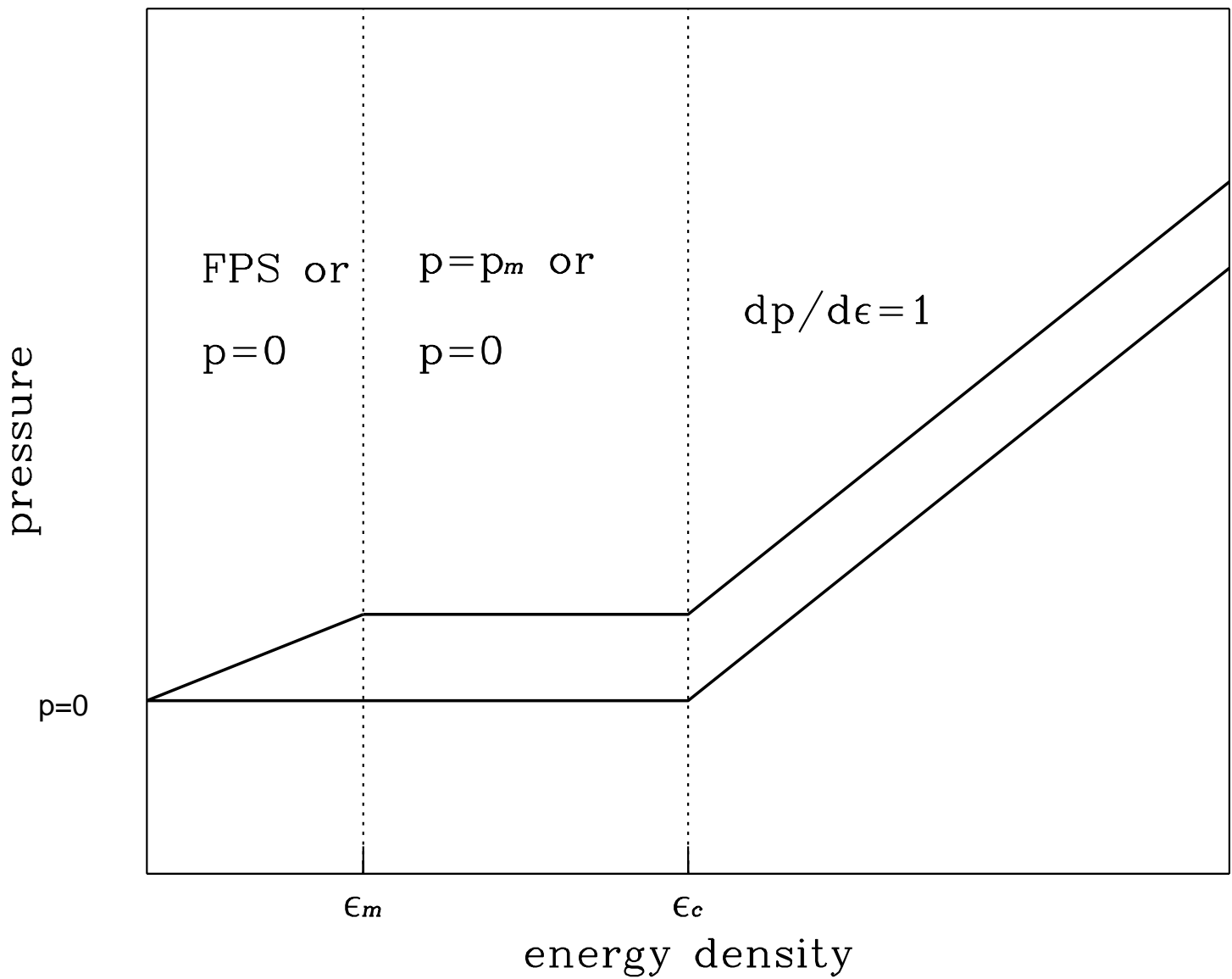
\begin{figure}[ht]
\caption{
\baselineskip 24 pt
Schematic representations of the minimum-period EOSs
(\protect\ref{eq:minimaleos1}) and (\protect\ref{eq:minimaleos}).  The
minimum-period EOS (\protect\ref{eq:minimaleos1}) does not match a
known low density EOS; the pressure vanishes for $\epsilon <
\epsilon_C$, and is at the causal limit with $dp/d\epsilon=1$ for
$\epsilon > \epsilon_C$.  The minimum-period EOS
(\protect\ref{eq:minimaleos}) matches the FPS EOS to a constant
pressure region at an energy density $\epsilon_m$. For $\epsilon >
\epsilon_C$ the EOS is at the causal limit
with $dp/d\epsilon=1$. Both axes are linear.}
\label{fig:LinearEos}
\end{figure}

Considering only assumptions (0) and (1), so that the EOS is
constrained only by causality, the EOS yielding the minimum period
stars for a given maximum spherical star mass $M_{sph}^{max}$ is
divided into two regions.  Above a density $\epsilon_C$ the
EOS is maximally stiff with $dp/d\epsilon=1$, while for $\epsilon \leq
\epsilon_C$ the EOS is maximally soft with vanishing pressure $p=0$: 
\begin{eqnarray}
p(\epsilon)&=&{\left\{
\begin{array}{ll} 
0  & \epsilon \leq \epsilon_C, \\ & \\
\epsilon-\epsilon_C & \epsilon \geq \epsilon_C.
\end{array}
\right.}
\label{eq:minimaleos1}
\end{eqnarray}
This EOS is depicted in Figure \ref{fig:LinearEos}. It gives a
stringent upper limit set by causality on the rotation of uniformly
rotating, gravitationally bound stars, independent of any specific
knowledge about the EOS for the matter composing the star. The energy
density $\epsilon_C$ is the single free parameter of this minimum-period
EOS.  By choosing $\epsilon_C$ one selects the maximum spherical star
mass $M_{sph}^{max}$ that the minimum-period EOS allows. Further, a one to
one correspondence exists between $\epsilon_C$ and other
characteristics of the minimum-period EOS, such as the maximum mass of
rotating stars.

\subsection{Minimum-Period EOS with FPS Low Density EOS}

Considering assumptions (0), (1) and (2) above, so that the EOS
matches the FPS EOS at low density, the minimum-period EOS is divided
into three regions.  Above a density $\epsilon_{C}$ the minimum-period EOS
is maximally stiff with $dp/d\epsilon = 1$. Between the matching
density to the FPS EOS $\epsilon_m$ and $\epsilon_C$ the EOS is
maximally soft with $dp/d\epsilon =0$. Below $\epsilon_m$ the
EOS is given by the FPS EOS. Explicitly,
\begin{eqnarray}
p(\epsilon)&=&{\left\{
\begin{array}{ll} 
p_{\rm FPS}(\epsilon) & \epsilon\leq \epsilon_m, \\ & \\
p_m  & \epsilon_m\leq \epsilon \leq \epsilon_C, \\ & \\
p_m + \epsilon-\epsilon_C & \epsilon \geq \epsilon_C.
\end{array}
\right.}
\label{eq:minimaleos}
\end{eqnarray}
This EOS is also depicted in Figure \ref{fig:LinearEos}.  The
maximally soft, constant pressure region corresponds to a first-order
phase transition for a single component system.  For a fixed low
density EOS and matching energy density $\epsilon_m$, the causal limit
energy density $\epsilon_C$ parameterizes the class of minimum-period
EOSs. Again, by choosing $\epsilon_C$ one selects the maximum
spherical star mass $M_{sph}^{max}$ that the minimum-period EOS
allows, and a one to one correspondence exists between
$\epsilon_C$ and other characteristics of the minimum-period EOS.
 
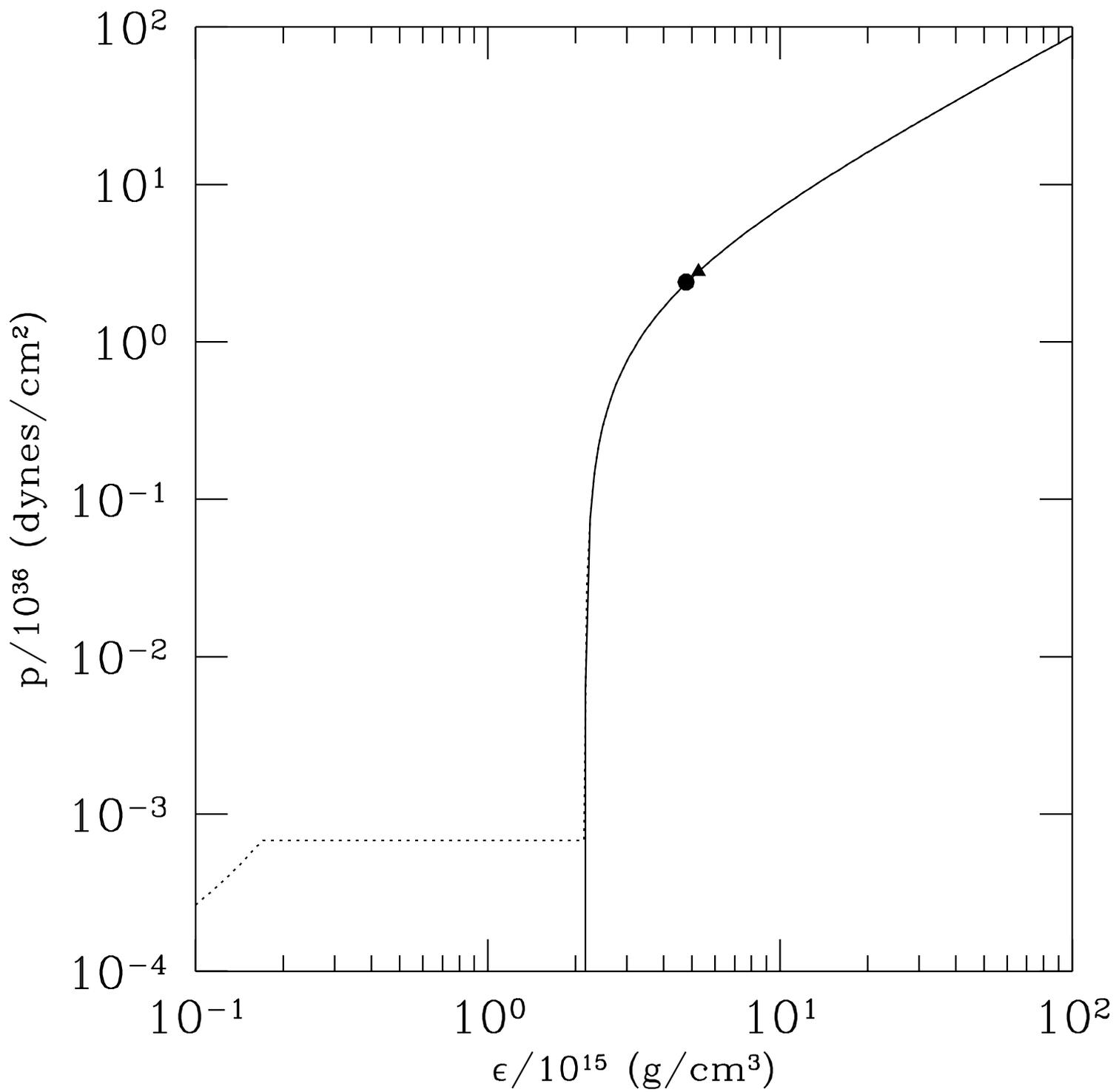
\begin{figure}[ht]
\caption{
\baselineskip 24 pt
Minimum period EOSs (\protect\ref{eq:minimaleos1}) and 
(\protect\ref{eq:minimaleos}) yielding a maximum-mass nonrotating star of
$M_{sph}^{max}=1.442\> {M}_\odot$. The solid curve is the minimum
period EOS (\protect\ref{eq:minimaleos1}), which does not match at low
density to a known EOS. The energy density at which the EOS becomes
causal is $\epsilon_C= 2.156\times 10^{15}{ g}/ { cm}^3$. The central
energy density of the minimum-period star is $ 4.778\times 10^{15}{
g}/{ cm}^3$ and is indicated by a filled circle on the solid curve.
The dotted curve is the minimum-period EOS
(\protect\ref{eq:minimaleos}), which matches to the FPS EOS at low
density. The energy density at which the EOS becomes causal is
$\epsilon_C= 2.157\times 10^{15}{ g}/ { cm}^3$. Above this
density the EOSs are nearly identical and the solid curve
overlaps the dashed curve. The central energy
density of the minimum-period star is $ 5.274\times 10^{15}{ g}/{
cm}^3$ and is indicated by a filled triangle on the dotted curve. The
matching number density is $n_m=0.1\>{ fm}^{-3}$. Both axes are
logarithmic.}
\label{fig:minimaleos}
\end{figure}

Figure \ref{fig:minimaleos} shows the minimum-period EOS
(\ref{eq:minimaleos1}) yielding a maximum mass spherical star with
$M_{sph}^{max}=1.442\>{M}_\odot$. The energy density at which the EOS
changes from zero pressure to $dp/d\epsilon=1$ is $\epsilon_C= 2.156
\times 10^{15}{ g}/{ cm}^3$, and the central energy density of the
minimum-period star is $4.778\times 10^{15}{ g}/{ cm}^3$. Also shown
in Figure \ref{fig:minimaleos} is the minimum-period EOS
(\ref{eq:minimaleos}) which matches to the FPS EOS at $n_m=0.1\>{
fm}^{-3}$ and yields a maximum mass spherical star with
$M_{sph}^{max}=1.442\>{M}_\odot$.  For this EOS $\epsilon_C=2.157
\times 10^{15}{ g}/ { cm}^3$ and the central energy density of the
minimum-period star is $5.274 \times 10^{15}{ g}/{ cm}^3$.  The
numerical work leading to the EOSs (\ref{eq:minimaleos1}) and
(\ref{eq:minimaleos}) is outlined in section
\ref{sec:numericalevidence}.   

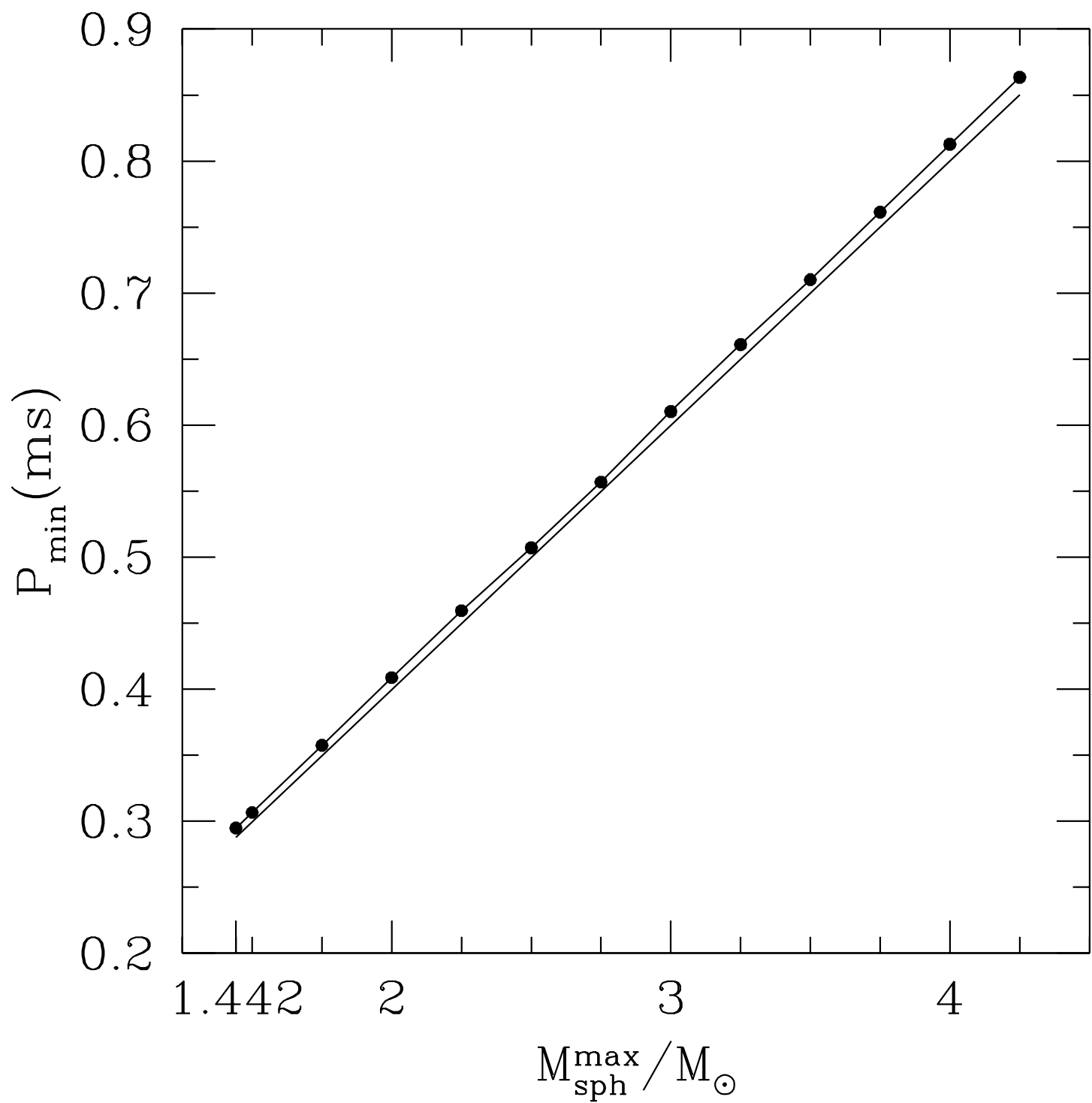
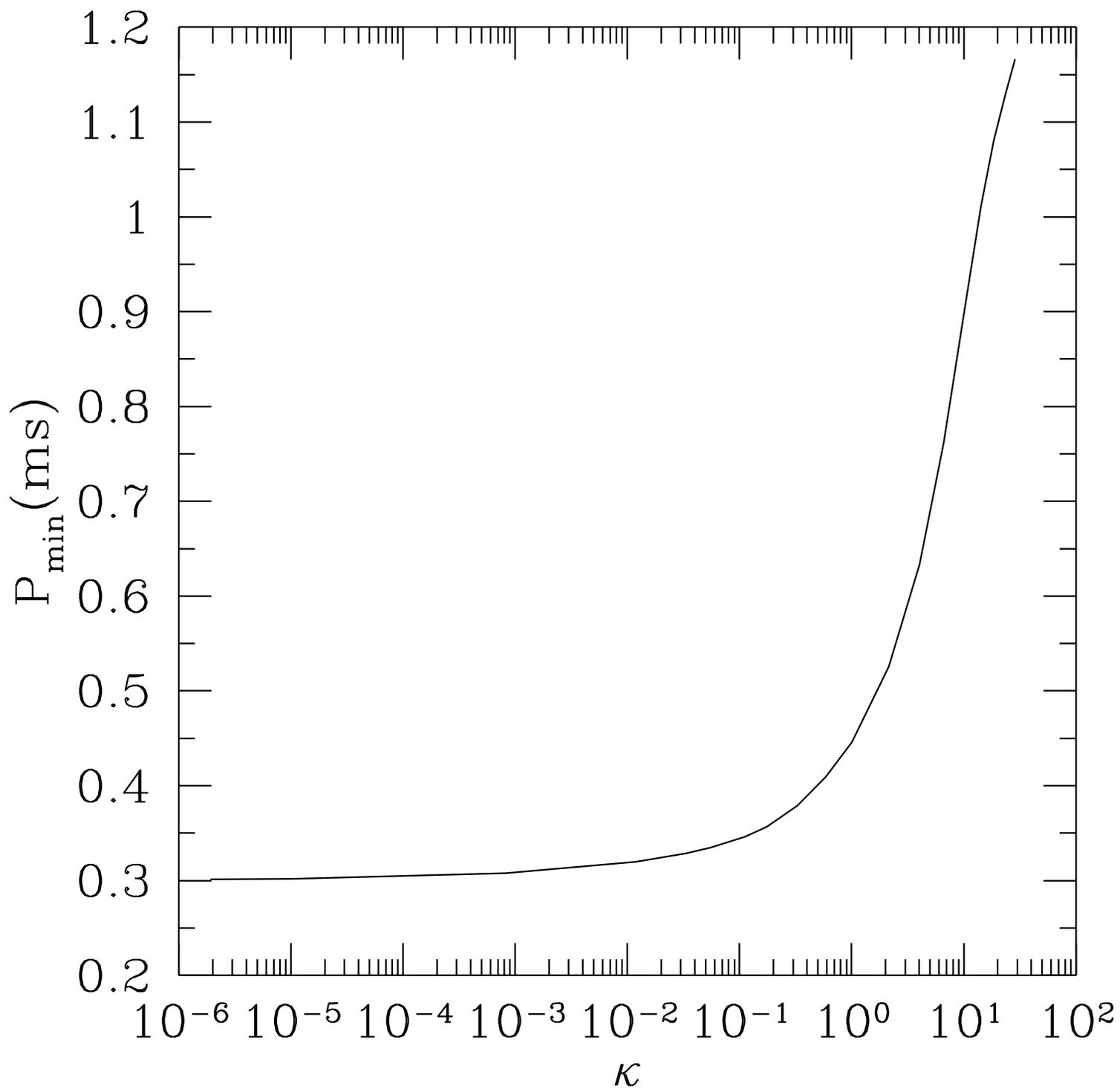
\begin{figure}[ht]
\caption{
\baselineskip 24 pt
The minimum period $P_{min}$ allowed for a rotating,
relativistic star as a function of the mass $M_{sph}^{max}$ of the
maximum-mass spherical star allowed by the EOS of the stellar
matter. The upper curve is constructed using the minimum-period EOS
(\protect\ref{eq:minimaleos}), which matches at low density to the FPS
EOS. The lower curve is constructed using the minimum-period EOS
(\protect\ref{eq:minimaleos1}), which does not match at low density to
a known low density EOS. Each curve divides the mass-period plane into
two regions. The region below a curve is not accessible to stars with
EOSs that satisfy the assumptions listed in section
\protect\ref{sec:eosconstraints}, and hence the region below the lower
curve is not accessible to stars composed of matter obeying causality.}
\label{fig:periodvsmass}
\end{figure}

Figure \ref{fig:periodvsmass} shows the minimum period $P_{min}$ of
gravitationally bound stars as a function of the maximum allowed mass
of nonrotating stars $M_{sph}^{max}$. The lower curve (which is a 
straight line) corresponds to
the minimum-period EOS (\ref{eq:minimaleos1}) satisfying assumptions
(0) and (1), while the upper curve corresponds to the minimum-period
EOS (\ref{eq:minimaleos}) matched to the FPS EOS at $n_m=0.1\>{
fm}^{-3}$.

The minimum-period EOS (\ref{eq:minimaleos1}) yields stars entirely at
the causal limit with a nonzero surface energy density $\epsilon_C$.
This surface energy density $\epsilon_C$ is the only free parameter
and the only dimensionful parameter. The family of maximally rotating
equilibrium stars yielded by the minimum-period EOS
(\ref{eq:minimaleos1}) with different $\epsilon_C$ are
characterized by this one dimensionful parameter; it follows that
all properties of the maximally rotating stars scale
according to their dimensions in gravitational units (with
$c=G=1$), $[\epsilon]= [M^{-2}]$, $[P]=[M]=[R]$.
Thus, the following relations hold between different
maximally rotating stars computed from minimum-period EOSs with
different $\epsilon_C$: 
\begin{eqnarray}
P&\propto &M_{sph}^{max} \propto R_{sph}^{max}, \\
\epsilon_{sph}^{max} &\propto& \frac{1}{\bigl ( M_{sph}^{max} \bigr )^2}, \\
M_{rot}^{max} &\propto& M_{sph}^{max}, \\
R_{rot}^{max} &\propto& R_{sph}^{max}, \\
\epsilon_{rot}^{max} &\propto& \epsilon_{sph}^{max}, 
\label{eq:relations}
\end{eqnarray}
where $R$ is the equatorial radius.   Numerically, we find the linear
relation $P_{\min}[ms] = 0.200 \ M_{sph}^{max}/M_\odot$, or
\begin{equation}
\frac{P_{\min}}{ ms} = 0.288 + 0.200 \ \left( 
   \frac{M_{sph}^{max}}{{ M}_\odot}-1.442 \right ).
\label{pmin1}
\end{equation}
(Note, however, that we have only considered the period of the
maximum-mass rotating model. In general the maximum-mass model is
distinct from the minimum-period model, but the difference in the
period between the two models is small. See section \ref{sec:accuracy}.)

The upper curve in Figure
\ref{fig:periodvsmass} corresponds to EOSs (\ref{eq:minimaleos}) which
match the FPS EOS at low density. This curve is almost linear, since
the matching number density $n_m=0.1 \ { fm}^{-3}$ is low enough that
the causal limit region of the EOS dominates the bulk properties of
the star. That is, a minimum-period star with a low matching density
mimics a star {\it entirely} at the causal limit and having a nonzero
surface energy equal to $\epsilon_C$. Hence the scaling relations
above are nearly exact for
minimum-period stars with low matching density. Numerically we
find for the  EOSs (\ref{eq:minimaleos}) 
\begin{equation}
\frac{P_{\min}}{ ms} = 0.295 + 0.203 \ \left( 
   \frac{M_{sph}^{max}}{{ M}_\odot}-1.442 \right ),
\end{equation}
which is linear to an accuracy better than 0.5 \%.  
For $M_{sph}^{max}=1.442\>{M}_\odot$ the minimum period is 0.29 ms.
The above formulae allow one to calculate without intensive numerical
computations the absolute minimum period if and when new observations
revise the current $M_{sph}^{max}=1.442 { M}_\odot$ limit.

\section{NUMERICAL EVIDENCE IN SUPPORT \\ OF THE MINIMUM-PERIOD EOS}
\label{sec:numericalevidence}

As noted above, our numerical results are consistent with the
expectation that the EOSs yielding the minimum-period stars are
maximally stiff at high density and maximally soft at low density.
Below we present the numerical evidence supporting our claim that the
EOSs (\ref{eq:minimaleos1}) and (\ref{eq:minimaleos}) yield the
maximum rotating stars, and detail how we searched the space of EOSs to
find the minimum-period EOS (\ref{eq:minimaleos}).

\subsection{Perturbations to the Minimum-Period EOS}
\label{sec:pert}

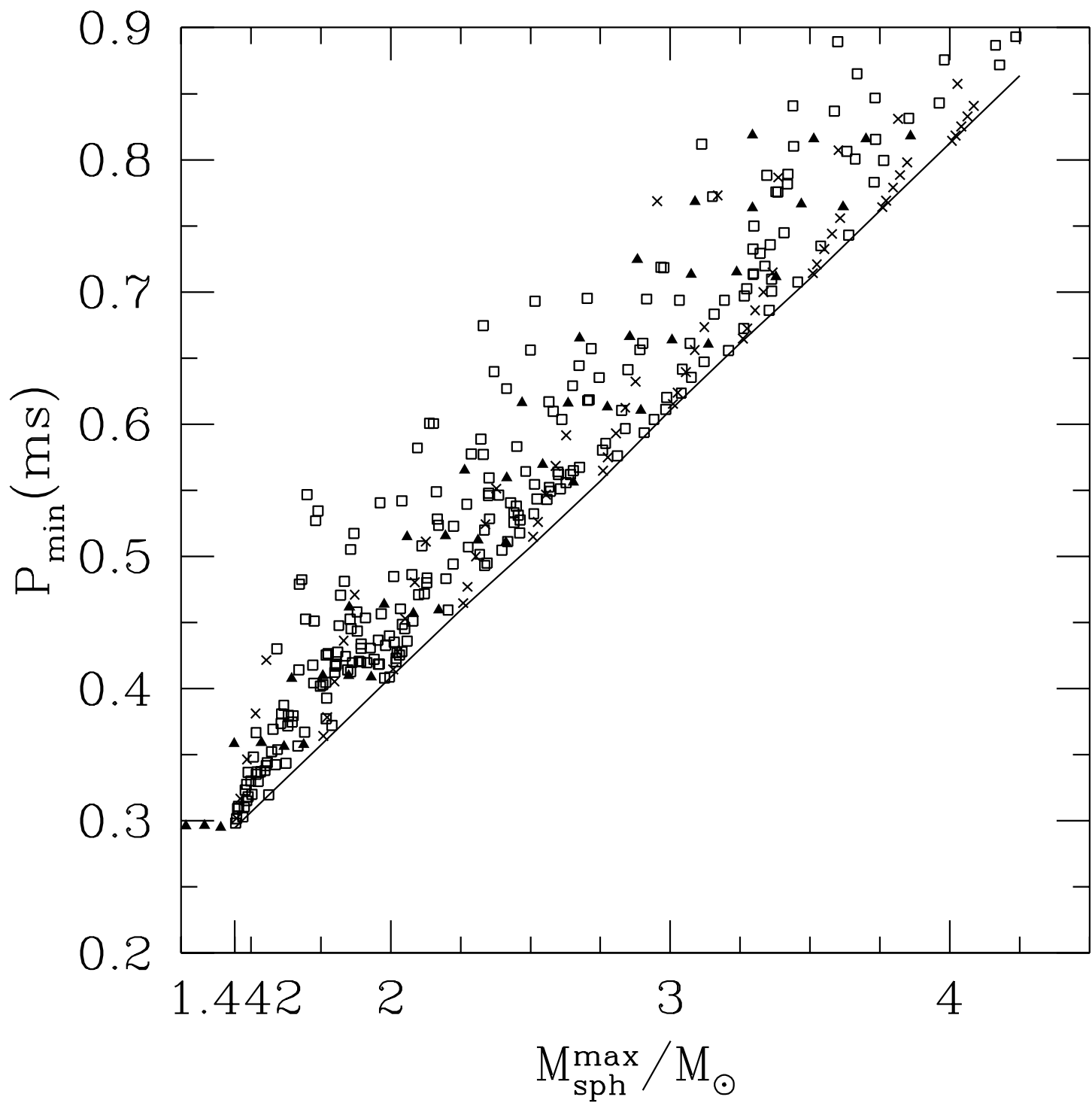
\begin{figure}[ht]
\caption{
\baselineskip 24 pt
Minimum period $P_{min}$ of rotating stars as a function of 
the allowed maximum mass $M^{max}_{sph}$ of spherical stars for EOSs
which are perturbations to the minimum-period EOS
(\protect\ref{eq:minimaleos}).  The squares, triangles, and crosses
represent EOSs obtained by different perturbations to the minimum
period EOS, which are detailed in section \protect\ref{sec:pert}.  The
solid curve represents the minimum-period EOS
(\protect\ref{eq:minimaleos}) and is the same as the upper curve in
Figure \protect\ref{fig:periodvsmass}. All of the points obtained
{}from perturbed minimum-period EOSs lie above the solid curve.}
\label{fig:pert}
\end{figure}

Evidence that the minimum-period EOS (\ref{eq:minimaleos}) yields
the fastest rotating stars is shown in Figure
\ref{fig:pert}. Each point in Figure \ref{fig:pert} represents an EOS
obtained by making a ``small'' perturbation to the minimum-period EOS
(\ref{eq:minimaleos}), and lies {\it above} the solid curve obtained
{}from the minimum-period EOS. For any particular $M_{sph}^{max}$ the
perturbed EOSs yield maximally rotating stars which rotate {\it
slower} than the maximally rotating stars yielded by the minimum-period
EOSs.

We considered three different types of perturbations to the minimum
period EOS. The first type of perturbed EOS considered is obtained
{}from the minimum-period EOS by adding a ``step function''
to the minimum-period EOS :
\begin{eqnarray}
p(\epsilon)&=&{\left\{
\begin{array}{ll} 
p_{\rm FPS}(\epsilon) & \epsilon \leq \epsilon_m, \\ & \\
p_m &  \epsilon_m \leq \epsilon \leq \epsilon_1, \\ & \\
p_m + \epsilon-\epsilon_1 &  \epsilon_1 \leq\epsilon \leq \epsilon_2, \\ & \\
p_m + \epsilon_2-\epsilon_1 & \epsilon_2 \leq\epsilon \leq \epsilon_C, \\ & \\
p_m + \epsilon_2-\epsilon_1 + \epsilon-\epsilon_C &
\epsilon \geq \epsilon_C. 
\\ & \\
\end{array}
\right.}
\label{eq:stepfunction}
\end{eqnarray}
By varying $\epsilon_1$ and $\epsilon_2$, as well as $\epsilon_C$, we sampled
a large number of perturbed minimum-period EOSs. For $\epsilon_2-\epsilon_1$
small, the minimum period for a given $M_{sph}^{max}$ allowed by the
perturbed EOS is slightly {\it larger} than that yielded by the minimum-period
EOS (\ref{eq:minimaleos}). The minimum periods obtained using the
``step function'' perturbations are indicated by the open squares
in Figure \ref{fig:pert}. For $\epsilon_2-\epsilon_1$ large, the minimum
periods yielded by the perturbed EOS are much larger and are not shown
in Figure \ref{fig:pert}. Thus a small perturbation away from the
minimum-period EOS (\ref{eq:minimaleos}) {\it increases} the minimum
period for a given $M_{sph}^{max}$.

The second type of perturbation to the minimum-period EOS we considered
is obtained by decreasing the slope $dp/d\epsilon$ away from unity
for the causal ($\epsilon \geq \epsilon_C$) part of the minimum-period
EOS. The filled triangles in Figure \ref{fig:pert} indicate the minimum
periods obtained by perturbing the minimum-period EOS in this way. Again,
the minimum period  increases when the minimum-period EOS 
(\ref{eq:minimaleos}) is perturbed. The third type of perturbation
we considered is obtained by increasing the slope $dp/d\epsilon$
away from zero for the constant pressure or $\epsilon_m \leq
\epsilon\leq\epsilon_C$ part of the minimum-period EOS. For 
$\epsilon \geq \epsilon_C$ the perturbed EOS again is at the causal limit
with $dp/d\epsilon=1$. The crosses in Figure \ref{fig:pert} indicate
the minimum periods obtained by perturbing the minimum-period EOS in
this way, and again the minimum period increases when the minimum
period EOS (\ref{eq:minimaleos}) is perturbed.

We have sampled a large number of perturbed minimum-period EOSs
which satisfy assumptions (0),(1), and (2) of section \ref{sec:eosconstraints},
and have found that the minimum period allowed by a perturbed EOS
for a given $M_{sph}^{max}$ is always greater than the minimum
period yielded by the minimum-period EOS (\ref{eq:minimaleos}).
These results, taken together with results
obtained from a two parameter ansatz EOS described in the next section,
strongly suggest that the minimum-period EOS
(\ref{eq:minimaleos}) is the EOS which provides the upper limit set by
causality for the rotation of gravitationally bound stars.

\subsection{Ansatz for Searching the Space of Equations of State}
\label{sec:ansatz}
 
To search the restricted space of EOSs which satisfy the assumptions 
in section \ref{sec:eosconstraints}, and to determine the class 
of EOSs which yield the minimum
period stars, we follow Glendenning (1992)
and adopt
a simple ansatz for the EOS.  For low number density, the energy
density and pressure are
\begin{eqnarray}
\epsilon(n)&=& \epsilon_{\rm FPS},\\
p(n)&=&p_{\rm FPS},
\end{eqnarray}
where $\epsilon_{\rm FPS}$ and $p_{\rm FPS}$
correspond to the FPS EOS. 
Above some number density $n_m$, where the FPS EOS is suspect,
the energy density and pressure are 
\begin{eqnarray}
\epsilon(n)&=&p_{m}\left\{\frac{\kappa}{\gamma-1}\left[
\left(\frac{n}{n_m}\right)^\gamma-\frac{n}{n_m}\right]
+\frac{n}{n_m}\frac{\epsilon_m}{p_m}+\left(1-\frac{n}{n_m}\right)
\left(\kappa-1\right)
\right\},
\label{eq:kgenergy} \\
p(n)&=&p_m \left\{ \kappa\left[\left(\frac{n}{n_m}\right)^\gamma
\right]+1\right\},
\label{eq:kgpressure}
\end{eqnarray}
where $\epsilon_m$ and $p_m$ are the energy density and pressure at
the matching number density $n_m$, which is typically $0.1 \ {
fm^{-3}}$.  The two dimensionless parameters $\kappa$ and $\gamma$
parameterize the restricted EOS space.  Assumptions (0) and (1) of
section \ref{sec:eosconstraints} require
\begin{eqnarray}
\gamma > &  1,\label{eq:gammarestriction} \\
\kappa > & 0, \label{eq:kapparestriction}\\
\kappa\gamma \leq & \frac{\epsilon_m}{p_m}+1.\label{eq:kgrestriction}
\end{eqnarray}
Given $\kappa$ and $\gamma$, we use the parameterized form
(\ref{eq:kgenergy}) and (\ref{eq:kgpressure}) for densities greater
than the matching density $n_m$, but only if the EOS remains causal.
If above some density $n_C$ the equation of state reaches the
causal limit, so that $\frac{dp}{d \epsilon} > 1$, then for $n > n_C$
the parameterized EOS is matched to an EOS which is always at the
causal limit. In this instance the energy density and pressure above
$n_C$ are
\begin{eqnarray}
\epsilon(n)&=&\epsilon_C-p_C+p(n),\label{eq:causalenergy}\\
p(n)&=&\frac{1}{2}\left\{ p_C-\epsilon_C+
\left(p_C+\epsilon_C\right)\left(
\frac{n}{n_C}\right)^2\right\},
\end{eqnarray}
where $\epsilon_C$ and $p_C$ are the energy density
(\ref{eq:kgenergy}) and pressure (\ref{eq:kgpressure}) evaluated at
$n_C$. Note that the parameterization of pressure as a function of
number density is arbitrary, since (\ref{eq:causalenergy}) assures
that $\frac{dp}{d\epsilon}=1$.  If $\kappa$ and $\gamma$ are chosen so
that the EOS remains causal then the parameterized form is used for all
densities $n\geq n_m$. Combining the different expressions for the
different ranges of number density the EOS is 
\begin{eqnarray}
\epsilon(n)&=&\left\{
\begin{array}{lll}
\epsilon_{\rm FPS}(n) & & n\leq n_m,\\ 
& & \\
p_{m}\left\{\frac{\kappa}{\gamma-1}\left[
\left(\frac{n}{n_m}\right)^\gamma- \right.\right. & \left. 
\frac{n}{n_m}\right]
  +\frac{n}{n_m}\frac{\epsilon_m}{p_m}  & \\
& \left. + \left(1-\frac{n}{n_m}\right)
\left(\kappa-1\right)
\right\}&  n_m\leq n \leq n_C, \\ 
& &\\
\epsilon_C-p_C+p(n) & & n \geq n_C,
\end{array}
\right.
\label{eq:wholeenergy} \\
p(n)&=&{\left\{
\begin{array}{ll}
p_{\rm FPS}(n) & n\leq n_m, \\ & \\
p_m \left\{ \kappa\left[\left(\frac{n}{n_m}\right)^\gamma
\right]+1\right\} & n_m\leq n \leq n_C, \\ & \\
\frac{1}{2}\left\{ p_C-\epsilon_C+
\left(p_C+\epsilon_C\right)\left(
\frac{n}{n_C}\right)^2\right\} & n \geq n_C,
\end{array}
\right.}
\label{eq:wholepressure}
\end{eqnarray}
where the $n \geq n_C$ case applies only if $n_C$ exists and is
finite.  Note that the EOS is continuous, although its first
derivative may be discontinuous at $n_m$.  We do not include a
separate expression for the possibility of a $p={ constant}$, 
one-component first-order phase transition region, since this may be
attained in the limit as $\kappa\rightarrow 0$.  One may search the
restricted EOS space using (\ref{eq:wholeenergy}) and
(\ref{eq:wholepressure}) by varying the parameters $\kappa$ and
$\gamma$.

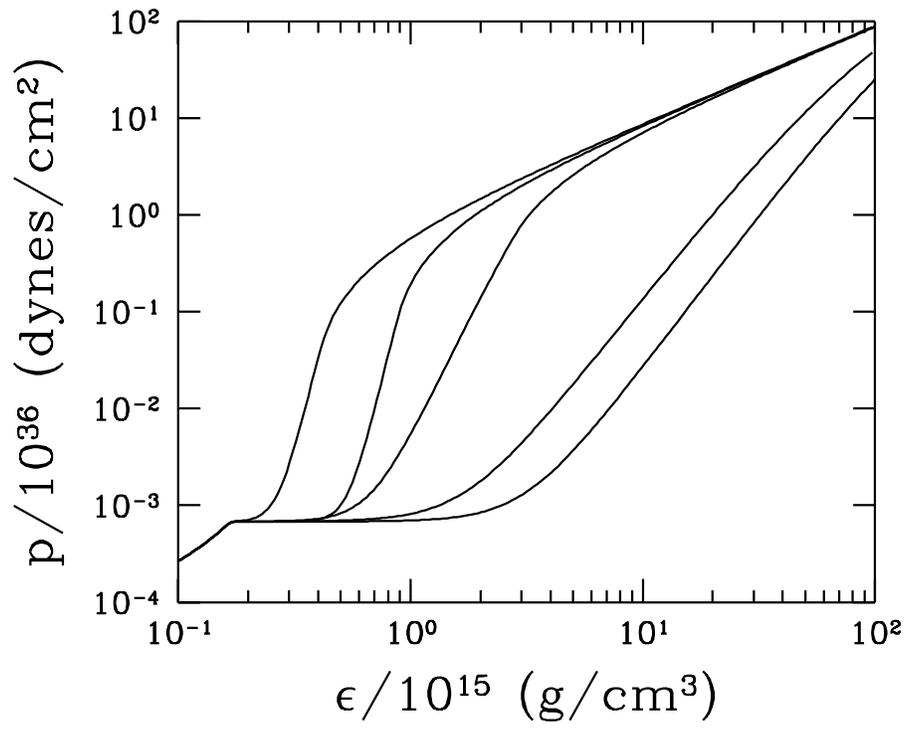
\begin{figure}[ht]
\caption{
\baselineskip 24 pt
Both plots show EOSs for different $\kappa$ and
$\gamma$ using the parameterization (\protect\ref{eq:wholeenergy}) and
(\protect\ref{eq:wholepressure}).  The $\kappa\gamma$ parameterized
EOS is fairly robust and can produce both stiff and soft
EOSs. The bottom plot shows a sequence of EOSs, all of which
yield $M_{sph}^{max}=1.442 M_\odot $. The period of the fastest
rotating star decreases from left to right as the
EOS approaches the minimum-period EOS (\protect\ref{eq:minimaleos}).
All axes are logarithmic.}
\label{fig:flexible}
\end{figure}

The two-parameter EOS (\ref{eq:wholeenergy}) and
(\ref{eq:wholepressure}) does not span the entire restricted EOS
space. Still, as Figure \ref{fig:flexible} shows, it is quite flexible,
and spans a representative subspace which includes both soft and stiff EOSs.
This two-parameter family of EOSs proved sufficient to 
identify the class of EOSs yielding the minimum-period stars.

\subsection{Searching for the Minimum-Period Star}
\label{sec:searchmethod}

We used a
fully relativistic, numerical code to compute rapidly rotating models
of compact stars and find the maximum-mass model for a given EOS.  
Stergioulas (Stergioulas \& Friedman 1995) wrote the
code following the Cook, Shapiro and Teukolsky (1994)
implementation of the Komatsu, Eriguchi
and Hachisu (1989) method. Koranda made changes which
substantially improved the speed of the code.

Using the flexible ansatz (\ref{eq:wholeenergy}) and
(\ref{eq:wholepressure}), we have searched for the class of EOSs which
yield the minimum-period stars. As mentioned in the introduction, we
are interested in the minimum period as a function of $M_{sph}^{max}$,
the mass of the maximum mass nonrotating (spherical) star. We first
used (\ref{eq:wholeenergy}) and (\ref{eq:wholepressure}) to search for
EOSs yielding particular values of $M_{sph}^{max}$.  Having determined
the $\kappa$ and $\gamma$ parameters for those EOSs yielding a
particular $M_{sph}^{max}$ , we then searched amongst those EOSs for
the one which yielded the rotating star with the minimum period  $P_{min}$. 
Figure \ref{fig:periodvskg} shows the results of our
search for $M_{sph}^{max}=1.442\> {M}_\odot$. Each point along the
curve defines an EOS which yields a maximum mass spherical star of
mass $M_{sph}^{max}=1.442\> {M}_\odot$. As one moves along the curve
towards smaller and smaller values of $\kappa$, {\it the minimum
period allowed by the EOS continues to decrease and tends
asymptotically to a minimum as} $\kappa\rightarrow 0$.

That the period tends only asymptotically to a minimum as
$\kappa\rightarrow 0$ can be understood by considering the EOSs
sampled as one moves along the curve in Figure
\ref{fig:periodvskg}. The bottom plot
of Figure \ref{fig:flexible} shows the EOSs sampled as one moves along
the curve in Figure \ref{fig:periodvskg}. As one moves to small
$\kappa$ the EOS is essentially unchanged, approaching the minimum
period EOS (\ref{eq:minimaleos}) with constant pressure from $p_m$ to
$p_C$. In the limit as $\kappa\rightarrow 0$ one obtains the minimum
period EOS (\ref{eq:minimaleos}) and the minimum period is given by
the upper curve in Figure \ref{fig:periodvsmass}.  This is true for all
reasonable $M_{sph}^{max}$.
 
\section{DEPENDENCE ON MATCHING\\ DENSITY}
\label{sec:matchdependence}

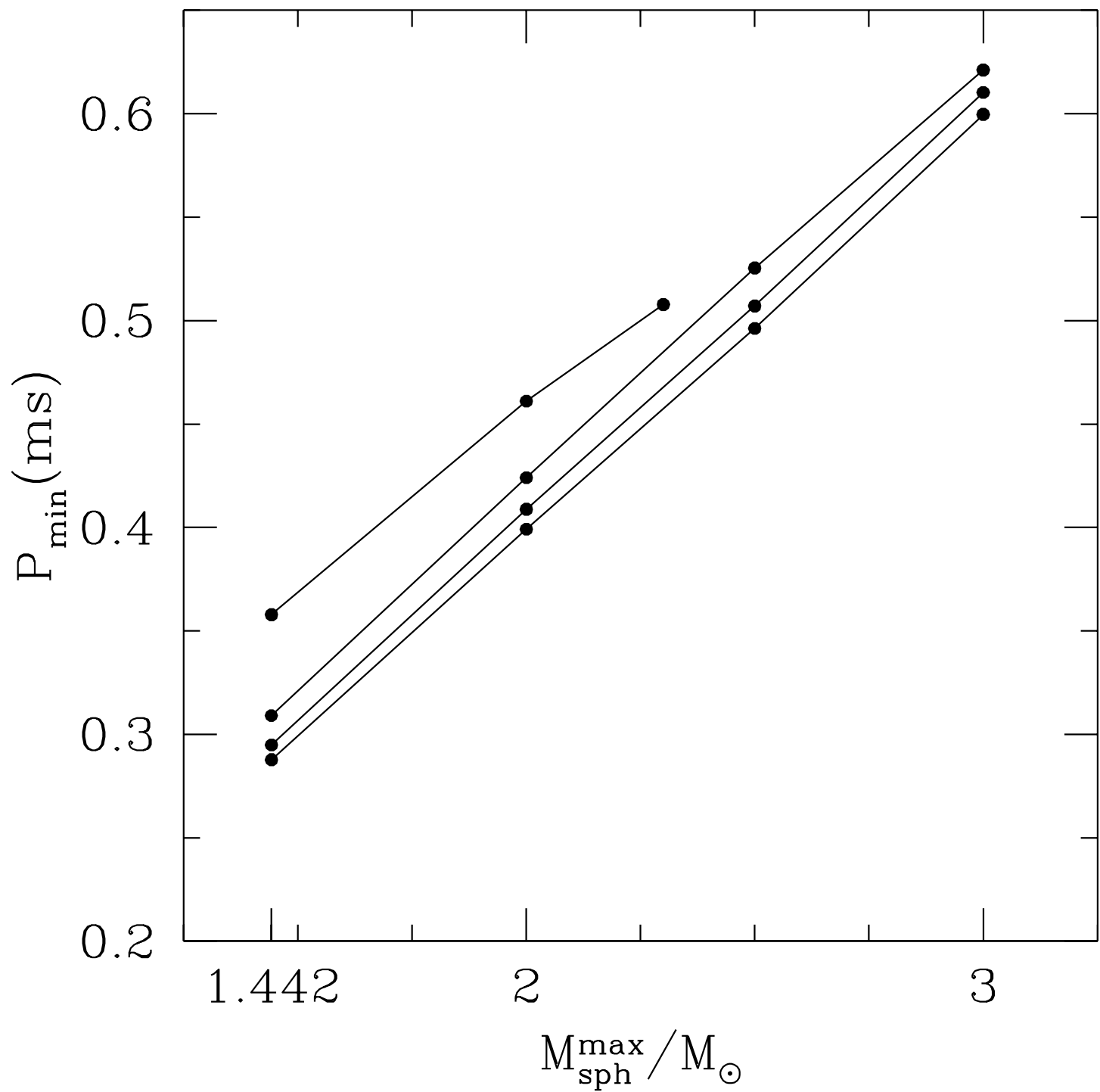
\begin{figure}[ht]
\caption{
\baselineskip 24 pt
The minimum period $P_{min}$ as a function of the mass 
$M_{sph}^{max}$ of the maximum-mass spherical star,
for different matching number densities $n_m$. From top to bottom
the curves represent $n_m=0.5, 0.25, 0.1$, and $0.0\ fm^{-3}$.
Note that for $n_m=0.5\ fm^{-3}$ the maximum-mass spherical model
cannot exceed $2.38\ M_\odot$.}
\label{fig:match}
\end{figure}

Our results are fairly insensitive to the matching density at which
the \linebreak[0]
min\-imum-period EOS (\ref{eq:minimaleos}) is matched to the FPS EOS for
low densities. This is because the matching density is low compared
to the central densities of the minimum-period stars, and as noted
above, the causality limited region of the EOS determines the bulk
properties of the minimum-period stars. Thus requiring that the star
have a crust of normal matter and satisfy assumption (2) of section
\ref{sec:eosconstraints} increases the
minimum period by less than $2.5\%$ for a matching density of
$n_m=0.1\ fm^{-3}$ and any $M_{sph}^{max}$.  We further increased the
matching density to $n_m=0.25 \ { fm}^{-3}$ (well beyond where one
trusts the FPS EOS) and the minimum period again increased by less
than 4.5 \% for any $M_{sph}^{max}$. This is shown in Figure
\ref{fig:match}. Our current understanding of the low density EOS
has a small effect on the minimum period determined only by the
causality constraint (1) of section \ref{sec:eosconstraints}.

\section{ACCURACY CHECK}
\label{sec:accuracy}

The numerical code used to construct rapidly rotating, fully
relativistic compact star models was checked for accuracy in an
extensive comparison to other codes (Eriguchi et al. 1996). The agreement
of the codes for physical quantities of specific models was 0.1 \%
to 0.01 \% or better, depending on the stiffness of the
EOS. Determining the maximum-mass model for a given EOS requires
computing a large number of models until one computes a model within
some tolerance of the true maximum mass and Kepler limit. Since the
mass vs. central energy density curve is approximately flat near the
maximum mass, we used a finely spaced grid and small tolerances (hence
a large number of computed models) 
to determine the maximum-mass model for a given EOS.

Still, for a given EOS the maximum-mass model is in general distinct
{}from the minimum-period model.  The intersection of the Kepler
limiting curve with the line of onset of axisymmetric instabilities in
a mass vs. central energy density plot defines the minimum-period
model (Stergioulas \& Friedman 1995; Cook et al. 1994).
The onset of axisymmetric
instabilities is defined by the relation
\begin{equation}
\Biggl( \frac{\partial M}{\partial \epsilon_{central}} \Biggr)_J =0,
\label{eq:axisymmetric}
\end{equation}
where $J$ is angular momentum. The central energy density at the
intersection may be higher or lower than the central energy density of
the maximum-mass model.  In general, however, the central energy
density, mass, period and other characteristics of the minimum-period
and maximum-mass models nearly coincide.  We have confirmed
that this is also true for the minimum-period EOSs
(\ref{eq:minimaleos1}) and (\ref{eq:minimaleos}).  Figure
\ref{fig:axiline} is a plot of mass vs. central energy density
for sequences of models with constant angular momentum constructed
using the minimum-period EOS (\ref{eq:minimaleos1}) and yielding
$M_{sph}^{max}=1.442 M_{\odot}$. Also shown is the Kepler limiting
curve and the axisymmetric instability line. The maximum-mass model has a
central energy density of $4.9\times 10^{15} g/cm^3$ and a period of 0.288 ms. 
The star having the maximum angular velocity for this equation of state has
a central energy density of $5.3\times 10^{15} g/cm^3$ and a period of 0.282 
ms, which is only $2\%$ less than the period of the maximum-mass model. 
Computing the period of the maximum-mass model, rather than the true minimum- 
period model, saves a tremendous amount of computing time and introduces an 
error not larger than 2 \%. 

\section{COMPARISON TO EARLIER WORK}
\label{sec:comparison}

Glendenning (1992) first estimated a causally limited 
period  for gravitationally bound stars. Rather than numerically 
computing rapidly rotating models, he computed
nonrotating models and estimated the minimum period for a given
EOS using the empirical formula 
\begin{equation}
\frac{P_{ min}}{ms} = 0.873 \left( {M_{sph}^{max}\over {M}_\odot} 
\right)^{-{1\over 2}}
\left({ R_{sph}^{max}\over { 10 \ km}} \right)^{3\over 2} .
\label{eq:empirical}
\end{equation} 
This empirical formula involves only the mass $M_{sph}^{max}$ and
radius $R_{sph}^{max}$ of the maximum mass nonrotating star for the
given EOS.  Haensel and Zdunik (1989) and Friedman {\it et al}
(Friedman, Ipser, \& Parker 1989) 
constructed this formula using realistic EOSs; it has
has an uncertainty of about 10 \% (the Haensel-Zdunik coefficient is 
more accurate and has in these units the value 0.83).  For a
1.442 ${M}_\odot$ maximum-mass nonrotating star Glendenning estimated
the absolute minimum period to be 0.32-0.33 ms, which is {\protect 10-14 \%}
larger than our computed value of 0.28 ms. The difference is as large
as the uncertainty of the empirical formula (\ref{eq:empirical}),
which is not surprising since (\ref{eq:empirical}) was constructed for a
a set of EOSs that is vastly different from the minimum-period EOSs.
Our numerical results yield an empirical formula for the class
of minimum-period EOSs which is fairly insensitive on the matching number
density $n_m$. For a matching density of $n_m=0.1 { fm}^{-3}$ the
empirical formula is
\begin{equation}
\frac{P_{ min}}{ms} = 
0.740 \left( {M_{sph}^{max}\over {M}_\odot} \right)^{-{1\over 2}}
\left({ R_{sph}^{max}\over { 10 \ km}} \right)^{3\over 2} ,
\label{eq:MinEmpirical}
\end{equation} 
with an approximate uncertainty of  0.3 \%. This formula is accurate to 
about 1\% for $0.0\leq n_m\leq 0.25\ fm^{-3}$.

Rather than identifying a single class of minimum-period EOSs,
Glendenning gave specific examples of minimum-period EOSs with and
without a constant pressure region. As we have demonstrated, there
is only one class of minimum-period EOSs, and each EOS in the class has
a constant pressure region.
We believe
that Glendenning missed this and proposed two classes of minimum
period EOSs because he used a modified Levenberg-Marquardt method
(Press et al. 1992) to minimize the function $f(M,P)=w_1(M-M_{sph}^{max})^2 +
 w_2 (P_{min})^2$, where $w_1$ and $w_2$ are weights. Our own trials
with this method showed that it is difficult simultaneously to obtain
high accuracy for both $M_{sph}^{max}$ and $P_{min}$.

\section{ABSOLUTE MAXIMUM MASS OF\\ UNIFORMLY ROTATING NEUTRON\\ STARS}
\label{sec:mass}

Among EOS satisfying assumptions (0-2) of section
\ref{sec:eosconstraints}, that which yields a spherical model of
maximum mass  has a simple and unique form, consisting of two parts. The
first part for low densities is a known low density EOS (earlier
authors used the BPS (Baym, Pethick, \& Sutherland 1971)
and Negele-Vautherin EOS).  The second
part, for densities greater than some
matching density $\epsilon_m$, is the causal limit EOS.  Rhoades and
Ruffini (1974) chose a particular $\epsilon_m$ and
computed a maximum mass of 3.2 ${M}_\odot$. Although 3.2 ${M}_\odot$
continues to be quoted as the theoretical maximum mass for neutron
stars, $M_{max}$ is sensitive to the matching energy density $\epsilon_m$
(as Hartle and Sabbadini (1977) pointed out) and
a more accurate statement of the upper limit set by causality on the
mass of compact spherical stars is  (Hartle 1978)
\begin{equation} 
M_{sph}^{max} \simeq 4.8 \ \Biggl ( \frac{2 \times 10^{14} \ { g/cm^3}}{
\epsilon_m} \Biggr )^{1/2} \ \ {M_\odot}.
\label{eq:Mmaxsph}
\end{equation}
This is in
sharp contrast to the class of minimum-period EOSs, where due to the
constant pressure region, the minimum period is insensitive to the
matching density. So 3.2 ${M}_\odot$ corresponds to a specific choice of
$\epsilon_m$ and is not the theoretical maximum mass of neutron stars,
which one needs to recompute whenever new certainty of the EOS allows
one to increase the matching density.
 
Friedman and Ipser (1992) assumed that the EOS limited by
causality yields the maximum mass for rotating stars as it does for
nonrotating stars.  They computed, using an independent code, the
maximum mass for different \linebreak[4]
match\-ing densities, assuming the BPS or
Negele-Vautherin EOS for low densities.  We also computed the maximum
mass of rotating
compact stars using the FPS EOS and our numerical code, which is based
on a different algorithm than the Friedman and Ipser code. Our results
are summarized as
\begin{equation}
M_{rot}^{max} \simeq 6.1 \ \Biggl ( \frac{2 \times 10^{14} \ { g/cm^3}}{
\epsilon_m} \Biggr )^{1/2} \ \ {M_\odot},
\label{eq:Mmax}
\end{equation}
which is 3 \% larger than the Friedman and Ipser result, and is within
the numerical accuracy of the code used by Friedman and Ipser.  Using
the FPS rather than the BPS EOS did not significantly affect the
maximum mass. In (\ref{eq:Mmax}), $2 \times 10^{14} \ { g/cm^3}$
is roughly the energy density up to which we trust the FPS EOS.  If
future research establishes the accuracy of FPS for higher energy
densities, then (\ref{eq:Mmax}) can be used to give an updated, lower
value for $M_{rot}^{max}$.


We acknowledge useful discussions with Bruce Allen, Lawrence Krauss,
Lee Lindblom, and Tanmay Vachaspati.  SK and NS acknowledge the
support of NSF Grant Nos. 91-05935 and 95-07740 while at the
University of Wisconsin-Milwaukee. SK acknowledges the support of Case
Western Reserve University and the DOE.

\newpage

\centerline{REFERENCES}

\parindent -0.5 in

Baym, G., Pethick, C., \& Sutherland, P. 1971, ApJ,  170, 229

Cook, G. B., Shapiro, S. L., \& Teukolsky, S. A. 1994, ApJ, 424, 823

Eriguchi, Y., Friedman, J. L., Gourgoulhon, E., Nozawa, T., \& 
Stergioulas, N. 1996, in preparation

Friedman, J. L., \& Ipser, J. R. 1987, ApJ, 314, 594

Friedman, J. L., \& Ipser, J. R. 1992, Phil. Trans. R. Soc. Lond., A340, 391

Friedman, J. L., Ipser, J. R., \& Parker, L. 1989, Phys. Rev. Lett., 62, 3015

Geroch, R., \& Lindblom, L. 1991,  Ann. Phys. 207, 394

Glendenning, N. K. 1992, Phys. Rev. D 46,4161 

Haensel, P., \& Zdunik, J. L. 1989, Nature, 340, 617

Hartle, J. B. 1978,  Phys. Repts., 46, 201

Hartle, J. B., \& Sabbadini A. G. 1977, ApJ, 213, 831

Komatsu, H., Eriguchi, Y., \& Hachisu, I. 1989, MNRAS, 239, 153

Pandharipande, V. R., \& Ravenhall, D. G. 1989,
in {\it Proceedings of the
NATO Advanced Research Workshop on Nuclear Matter and Heavy Ion Collisions, 
Les Houches, 1989}, ed.  M. Soyeur {\it et al.} (New York: Plenum),
103

Press, W. H., Flannery, B. P., Teukolsky, S. A., 
\& Vetterling, W. T. 1992, Numerical Recipes in C,
2nd edition, (Cambridge: Cambridge Univ. Press) 

Ravenhall, D. G. 1995, personal communication

Rhoades, C. E. Jr., \& Ruffini R. 1974, Phys. Rev. Lett., 32, 324

Stergioulas, N., \&  Friedman, J. L. 1995, ApJ, 444, 306

Taylor, J. H., \& Weisberg, J. M. 1989, ApJ, 345, 434

\newpage


\begin{figure}[ht]
\caption{
\baselineskip 24 pt
Minimum period allowed by EOSs given by
(\protect\ref{eq:wholeenergy}) and (\protect\ref{eq:wholepressure})
for different values of the parameter $\kappa$. For given $\kappa$, 
the parameter $\gamma$ is determined so that all EOSs sampled as one
moves along the curve yield a maximum-mass nonrotating star with
$M_{sph}^{max}=1.442\>{M}_\odot$.  As $\kappa\rightarrow 0$ the period
tends asymptotically to its minimum value. }
\label{fig:periodvskg}
\end{figure}

\begin{figure}[ht]
\caption{
\baselineskip 24 pt
Mass vs. central energy density for sequences of models
with constant angular momentum, constructed using the minimum-period
EOS (\protect\ref{eq:minimaleos1}) with $M_{sph}^{max}=1.442 \> M_\odot$. 
The solid curves are sequences of
models with constant angular momentum, the bottom one corresponding to
nonrotating models.  The dashed curve is the sequence of models rotating at 
the Kepler limit while the dotted curve is the axisymmetric instability line 
defined by (\protect\ref{eq:axisymmetric}). The maximum-mass model  is
distinct from the model with maximum angular velocity
$\Omega_{max}$. The period of the maximum-mass model
is only 2$\%$ less than the period of the fastest rotating model.}
\label{fig:axiline}
\end{figure}

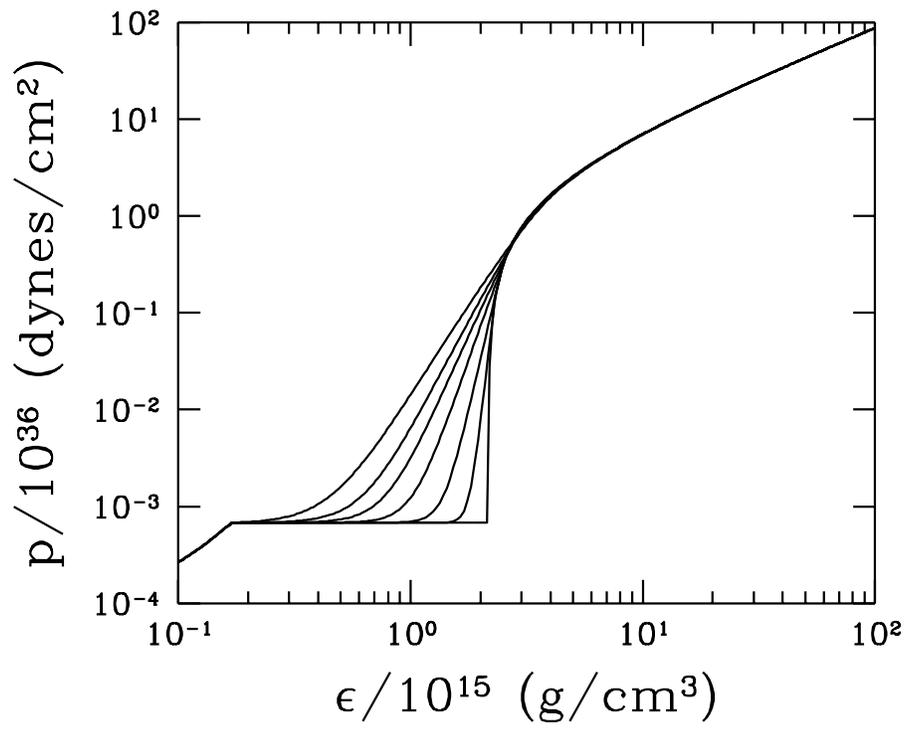
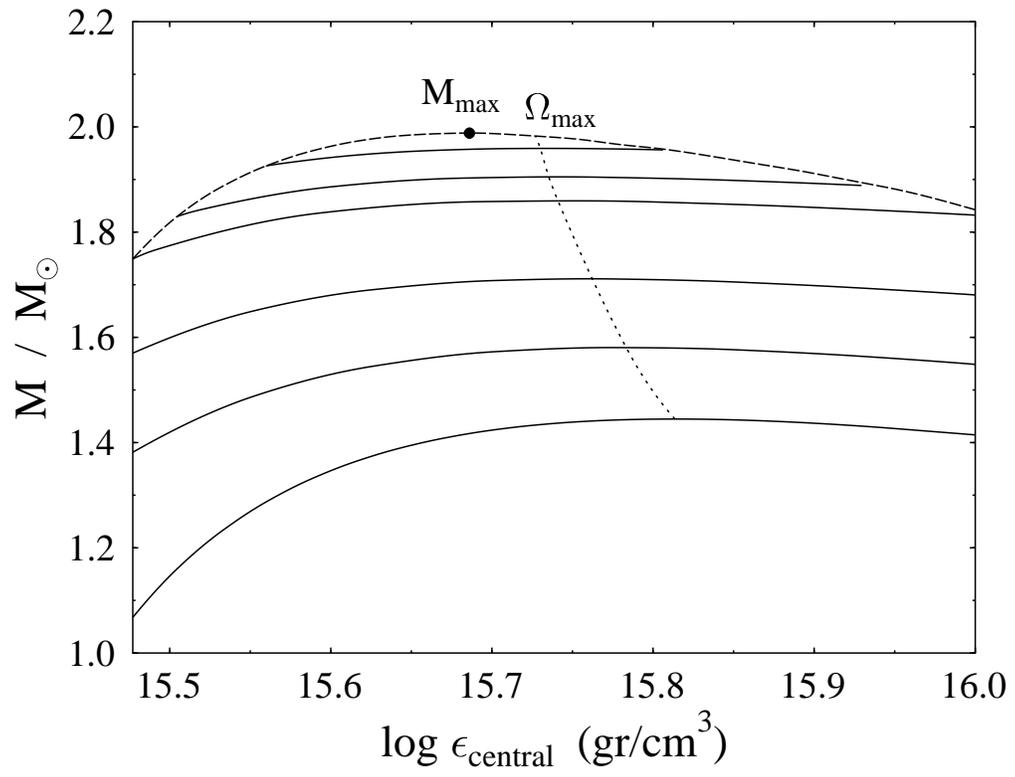
\end{document}